\documentclass[a4paper,12pt,dvipsnames,usernames]{article}
\pdfoutput=1

\usepackage{aas_macros}

\usepackage{jcappub} 

\usepackage[utf8]{inputenc}
\usepackage{graphicx}
\usepackage{amsmath}
\usepackage{amssymb}
\usepackage{bm}
\usepackage{paralist,array}
\usepackage{units} 
\graphicspath{{fig/}{./Plots/}} 
\usepackage{slashed}
\usepackage{todonotes}
\usepackage[parfill]{parskip}
\usepackage{longtable}
\usepackage{subfig}

\graphicspath{{plots/}}

\title{Probing intergalactic intergalactic magnetic fields with LOFAR LoTSS DR2 data}

\author[1]{Kyrylo Bondarenko,}
 \emailAdd{kyrylo.bondarenko@su.se}
\affiliation[1]{
Nordita, KTH Royal Institute of Technology and Stockholm University, Hannes Alfv\'ens v\"ag
12, 10691 Stockholm, Sweden}

\author[2]{Alexey Boyarsky,}
 \emailAdd{boyarsky@lorentz.leidenuniv.nl}
\affiliation[2]{%
 Institute Lorentz, Leiden University, Niels Bohrweg 2, Leiden, NL-2333 CA, the Netherlands
}

\author[3,4]{Anastasia Sokolenko,}
\emailAdd{sokolenko@kicp.uchicago.edu}
\affiliation[3]{
Theoretical Astrophysics Department, Fermi National Accelerator Laboratory, Batavia, Illinois, 60510, USA
}
\affiliation[4]{
Kavli Institute for Cosmological Physics, The University of Chicago, Chicago, IL 60637, USA
}

\author[5]{Ievgen Vovk}
 \emailAdd{vovk@icrr.u-tokyo.ac.jp}
\affiliation[5]{
Institute for Cosmic Ray Research, The University of Tokyo, 5-1-5 Kashiwa-no-Ha, Kashiwa City, Chiba, 277-8582, Japan}

\begin{document}

\abstract{
We use Faraday rotation measurements from the latest catalog LoTSS DR2 from LOFAR to probe intergalactic magnetic fields. To identify the extragalactic component of the observed rotation measure (RM) we use two different techniques: residual rotation measure (RRM) and close radio pairs. For the RRM approach, we conclude that, despite smaller measurement errors in the LOFAR data, robust and conservative treatment of the systematic uncertainties in the Galactic contribution to RM results in the constraint on a homogeneous volume-filling magnetic field at the level 2.4 nG, slightly weaker than previous constraints from NVSS data, and does not allow to probe the presence of over-magnetized bubbles predicted by the AGN feedback model of the IllustrisTNG.
Analyzing close radio pairs we found that in only 0.5\%  of our mock realizations of observed data, the expected contribution from the over-magnetized bubbles does not exceed LoTSS DR2 data. 
% Analyzing close radio pairs we found that in only 158 of our mock realizations of observed data from 30000 the expected contribution from the over-magnetized bubbles does not exceed LoTSS DR2 data. 
}

\maketitle

\section{Introduction}

Primordial magnetic fields (PMF), if detected, would be a new pillar of modern cosmology. By studying the properties of PMF one could expect to understand the state of our Universe at the epochs earlier than BBN, currently our earliest cosmological probe~\cite{Grasso:2000wj,Battaner:2000kf,Giovannini:2003yn,2012SSRv..166....1R,2012SSRv..166...37W,Durrer:2013pga,Subramanian:2015lua}. Therefore, the task of developing the observational methods that allow to measure of PMF is of great importance. In this paper, we discuss two techniques based on observations of Faraday rotation.

The Faraday rotation technique allows to probe of magnetic fields in dense regions such as galaxies and galaxy clusters, as well as to constrain the magnetic field in the intergalactic medium~\cite{Beck:2013bxa,2015A&ARv..24....4B,Ferrari:2008jr,2010A&A...513A..30B,2012A&ARv..20...54F,Blasi:1999hu,Pshirkov:2015tua,Aramburo-Garcia:2022ywn}. Probing the magnetic field outside collapsed structures is a challenge. The Rotation Measure (RM) probes the product of the parallel component of the magnetic field along the line of sight, $B_{||}$, and the electron density, $n_e$, and is defined as 
\begin{equation} 
   \text{RM} = \frac{e^3}{2\pi m_e^2}\int \frac{n_e B_{\parallel}}{(1+z)^2} \frac{{\rm d}\ell}{{\rm d}z} {\rm d}z,
   \label{eq:FRMeq}
\end{equation}
where $m_e$ is the electron mass, $e$ is the electron charge, and $z$ is a redshift and $\ell$ is a physical distance along the line of sight.

In the last years, new RM data with much higher precision from LOFAR has become available, LoTSS survey~\cite{OSullivan:2023eub}. The previous major survey NVSS~\cite{Condon1998,Taylor2009} made by VLA telescope was based on the analysis of only two radio bandwidths and was subject to significant measurement uncertainties and may be subject to a wrapping uncertainty, which is not the case for LoTSS. So, this new data might potentially improve constraints on the intergalactic magnetic fields (IGMF).

The RM consists of three contributions: (i) the local environment near the host galaxy, (ii) the intergalactic medium (IGM), and (iii) our own Galaxy. There is quite extensive literature that studies Galactic contribution, based on the measurement of extragalactic RM sources and additional observations (e.g. free free emission)~\cite{Frick:2000fd,Johnston-Hollitt:2004dyq,Dineen:2004ys,Xu:2005rb,2012A&A...542A..93O,Oppermann:2014cua,2020A&A...633A.150H,2022A&A...657A..43H}. Therefore, one can divide RM into Galactic contribution, the so-called Galactic Rotation Measure (GRM) and extragalactic contribution, the Residual Rotation Measure (RRM). Derived RRM provides a constraint on the volume-filling intergalactic magnetic fields (IGMF)~\cite{1994RPPh...57..325K,Blasi:1999hu,Neronov:2013lta,Pshirkov:2015tua,Aramburo-Garcia:2022ywn}. Using the NVSS survey, one can put an upper bound on the IGMF at the level of $B \sim 2 \cdot 10^{-9}$~G~\cite{Pshirkov:2015tua,Aramburo-Garcia:2022ywn}. 

To estimate RRM for LoTSS data, there are two recent GRM maps available:~\cite{2020A&A...633A.150H} (we will call it GRM 2019 throughout the paper) and~\cite{2022A&A...657A..43H} (we will reference to it as GRM 2020). The key difference between them is that GRM 2020 uses the LoTSS dataset to build the galactic RM model, while GRM 2019 is created without measurements from LoTSS. Therefore, we will use both maps in this paper, to check the consistency between them or see possible differences.

The RM contribution from the IGM can be divided into the contribution from the volume-filling magnetic field (e.g. primordial magnetic fields), the contribution that comes from the activity of galaxies, such as the galactic outflows that eject magnetized matter from the galaxies, and ``pollute'' the IGM far beyond the galaxies. 

In~\cite{Garcia:2020kxm}, utilizing the state-of-the-art cosmological simulation, the IllustrisTNG simulation, it was demonstrated that processes during galaxy evolution might significantly pollute the IGM with the magnetic fields created inside galaxies. That results in the formation of the so-called over-magnetized ``bubbles'' around collapsed structures. These over-magnetized ``bubbles'' might extend to tens of megaparsecs in size. While the electron number density in these magnetic bubbles might be close to the average electron number density in the Universe, the value of the magnetic field might be several orders of magnitude larger than the volume-filling magnetic field. Even though the IllustrisTNG utilizes a realistic model of galaxy formation and baryonic feedback, it is not guaranteed that magnetic bubbles exist. 

In this paper, we would like to probe the IllustrisTNG model with LoTSS data as well as reanalyze the constraints on the volume-filling magnetic field. In particular, we will use two approaches, utilizing the residual rotation measure and using the method of close radio pairs~\cite{Vernstrom:2019gjr,OSullivan:2020pll,Pomakov:2022cem}. The principal idea is that for a closed pair of radio sources, the difference between their RM should approximately cancel our contribution from the Galaxy, leaving part of RM from the IGM in between the two sources.

The structure of the paper is as follows: in Sec.~\ref{sec:IllustrisTNG} we describe the properties of IllustrisTNG simulation including magnetic bubbles. In Sec.~\ref{sec:LoTSS} we briefly remind the reader of the main properties of the LoTSS DR2 dataset. Next, in Secs.~\ref{sec:RRM} and~\ref{sec:radio-pairs} we perform our main analysis using the method of residual rotation measures and close radio pairs correspondingly. Finally, we present our conclusions in Sec.~\ref{sec:conclusions}.

\section{IllustrisTNG simulations}
\label{sec:IllustrisTNG}
%%%%%%%%%%%%%%%%%%%%%%%%%%%%%%%%%%%%%%%%%%%%%%%%

The IllustrisTNG simulation suite constraint a number of different gravo-mag\-ne\-to\-hyd\-ro\-dy\-na\-mic simulations~~\citep{nelson18,springel18,pillepich18,naiman18,marinacci18}, based on the moving-mesh \textsc{Arepo} code \citep{springel2010MNRAS.401..791S}. It solves the system of equations for self-gravity and ideal magnetohydrodynamics~\citep{2011MNRAS.418.1392P,2013MNRAS.432..176P}. In this work, we use a high-resolution simulation box TNG100-1 with the size of $110$~cMpc. It constraints $1820^3$ dark matter particles and the same amount of gas cells with masses of $m_{\text{DM}} = 7.5 \times 10^6~{\rm M}_{\odot}$ and $m_{\text{bar}} = 1.4 \times 10^6~{\rm M}_{\odot}$ correspondingly. The IllustrisTNG has a comprehensive galaxy formation model, that includes the formation and evolution of supermassive black holes and their activity via thermal heating of the gas and high-velocity kinetic winds, see~\cite{Garcia:2020kxm} for more details.

The initial conditions for the magnetic field in the simulation were chosen to be a homogeneous magnetic field with strength $B_0 = 10^{-14}$\,cG (comoving Gauss). Magnetic fields with such superhorizon correlation length could be produced, for instance, during the inflation, see~\cite{Durrer:2013pga,Subramanian:2015lua} for a review.

\subsection{Magnetic bubbles in IllustrisTNG simulations}
\label{sec:bubbles-small-B0}

It was found~\citep{Garcia:2020kxm,Garcia:2021cgu,Bondarenko:2021fnn} that the IllustrisTNG simulation predicts the existence of magnetized bubbles which are formed a rather low redshifts $z \lesssim 2$ because of baryonic feedback from galaxies, mostly in the form of ejection material from AGNs. These bubbles have a magnitude of magnetic field of a few orders larger than the initial seed field in the simulation and could occupy cosmological regions with diameters of tens of Mpc. Varying value of the initial magnetic field in the simulation by many orders of magnitude, \cite{Garcia:2020kxm,Garcia:2021cgu} showed that properties of magnetic bubbles are largely independent of the initial magnetic field.

The distribution of magnetic field in the simulation volume at low redshift consists of two separate brunches~\cite{Aramburo-Garcia:2022ywn}: one that corresponds to the initial seed magnetic field with magnitude of $\sim 10^{-14}$~cG and the second one that represents the magnetic bubbles with characteristic field values $\sim 10^{-9}$~cG. The magnetic field in both branches scales with baryonic number density according to adiabatic contraction law, $B\propto n^{2/3}$. Following~\cite{Aramburo-Garcia:2022ywn}, we divide the simulation volume into two regions: a region with magnetic field $|B|>10^{-12}$~cG that correspond to the volume influenced by galactic outflows (the region with the magnetic bubble), and all other volume, that is mostly occupied by the magnetic field that correspond to the evolved seed magnetic field, which we call region with primordial magnetic field.

\section{LoTSS data}
\label{sec:LoTSS}

\begin{figure}[h!]
    \centering
    \includegraphics[width=0.95\linewidth,trim={0 0 0 1cm},clip]{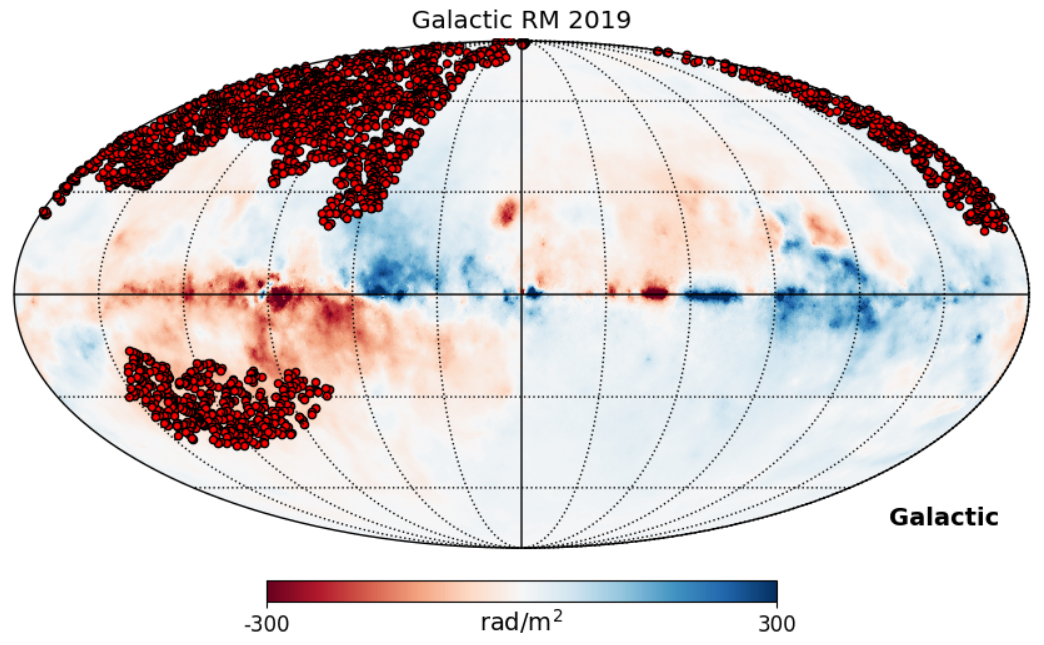}
    \caption{Visualization of positions of objects from LoTSS survey~\cite{OSullivan:2023eub} on the galactic plane (red points) over Galactic rotation measure map~\cite{2020A&A...633A.150H}.}
    \label{fig:LOFARdataGalactic} 
\end{figure}

In this work, we use the second data release (DR2) from the LOFAR Two-metre Sky
Survey (LoTSS)~\cite{OSullivan:2023eub}. It contains 2461 extragalactic high-precision RM values, that are collected from 5720~deg$^2$ of the sky, see Fig.~\ref{fig:LOFARdataGalactic} for the visualization of objects' location in galactic coordinates. We see that observations are done in two unconnected regions in the northern and southern hemispheres with 2039 and 422 objects respectively.

Polarization properties of each source were obtained using RM synthesis technique~\cite{1966MNRAS.133...67B,Brentjens:2005zc} from Stokes Q and U channel images, that have angular resolution of $20''$ and were measured with the frequency range from 120 MHz to 168 MHz with a channel bandwidth of 97.6 kHz. The resulting catalog has a median degree of polarization of 1.8\% (from 0.05\% up to 31\%). Comparison of sources from this catalog that are also present in NVSS survey~\citep{Condon1998,Taylor2009} measured at frequency $1.4$~GHz shows the good agreement between derived RM values, which means that LoTSS objects have minimal amounts of Faraday complexity and are good sample for extragalactic RM studies.

\begin{figure}
    \centering
    \includegraphics[width=0.75\textwidth]{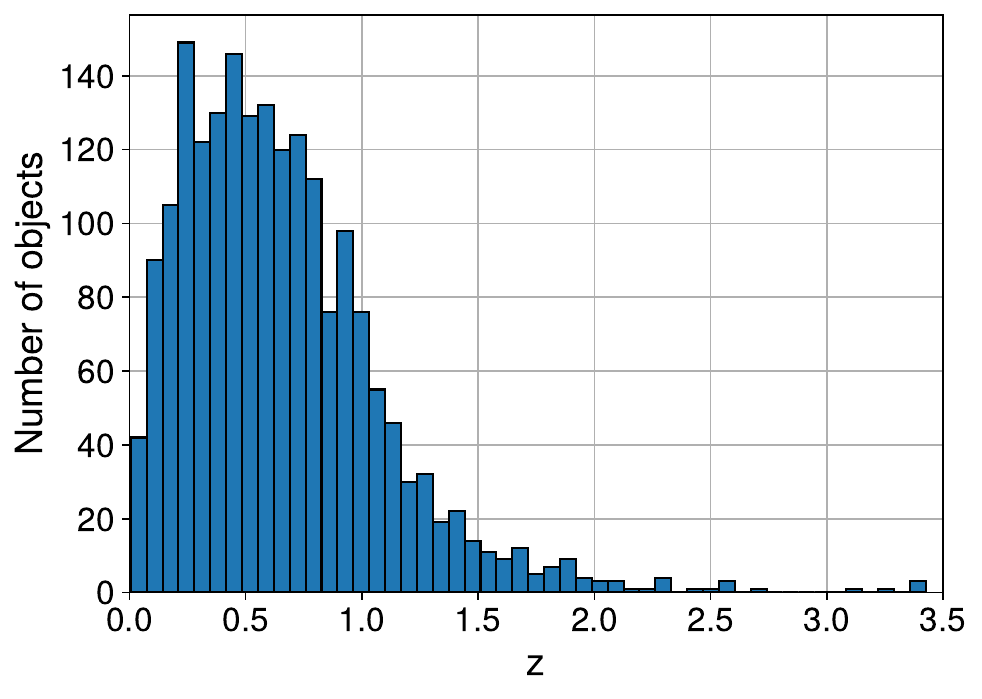}
    \caption{Redshift distribution of objects from LoTSS catalog with identified photometric or spectroscopic redshifts.}
    \label{fig:LOFAR-z-distribution}
\end{figure}

Radio observations usually do not contain specific features in the spectrum to allow identifying the redshift of the source. Therefore, matching with optical observation was made within the LOFAR Galaxy Zoo effort~\cite{Williams:2018dss} within the Surveys and Magnetism Key Science Project
teams. Each source was classified by five different astronomers, which resulted in the identification of a host galaxy for 2168 of the 2461 polarized components (88\% of the total catalog). For most of these host galaxies, photometric or spectroscopic redshifts are available, which allows to assign redshifts for 79\% of objects in the LoTSS catalog, see Fig.~\ref{fig:LOFAR-z-distribution} with the distribution of objects by redshift.

\section{Residual rotation measure}
\label{sec:RRM}
\subsection{Galactic RM maps}
\label{subsec:grm}

The residual rotation measure (RRM) is an extragalactic contribution to the observed total RM. This is the most direct probe of extragalactic RM. However, to utilize it one should know well enough the galactic rotation measure (GRM) contribution.

There are two recent models of GRM~\cite{2020A&A...633A.150H, 2022A&A...657A..43H}, which we refer in this paper as GRM 2019~\cite{2020A&A...633A.150H} and GRM 2020~\cite{2022A&A...657A..43H}.
Both are created with a similar technique of the Wiener filter~\cite{Ensslin:2008iu}. The main difference between them is that GRM 2019 uses a dataset of 41,632 extragalactic radio sources with measured RM and a map of free free emission from the cosmic microwave background (CMB) foreground studies~\cite{Planck:2015mvg}.\footnote{Free free emission as bremsstrahlung radiation created by scattering of free electrons on free protons. Its intensity is proportional to the square of electron number density along the line of sight, which is parametrically close to the rotation measure, see~\cite{2020A&A...633A.150H} for more discussions.} The GRM 2020 map uses a larger catalog of 55,190 extragalactic sources (including LoTSS data studied in this paper), does not use the free free emission data, and produces a resulting map with twice higher angular resolution compared to the GRM 2019. 
Below we discuss the differences between these maps and implications for the conclusions we might derive from the LoTSS RRM.

\subsection{RRM and magnetic bubbles}
\label{subsec:rrm}

\begin{figure}[t]
    \centering
    \includegraphics[width=0.75\linewidth]{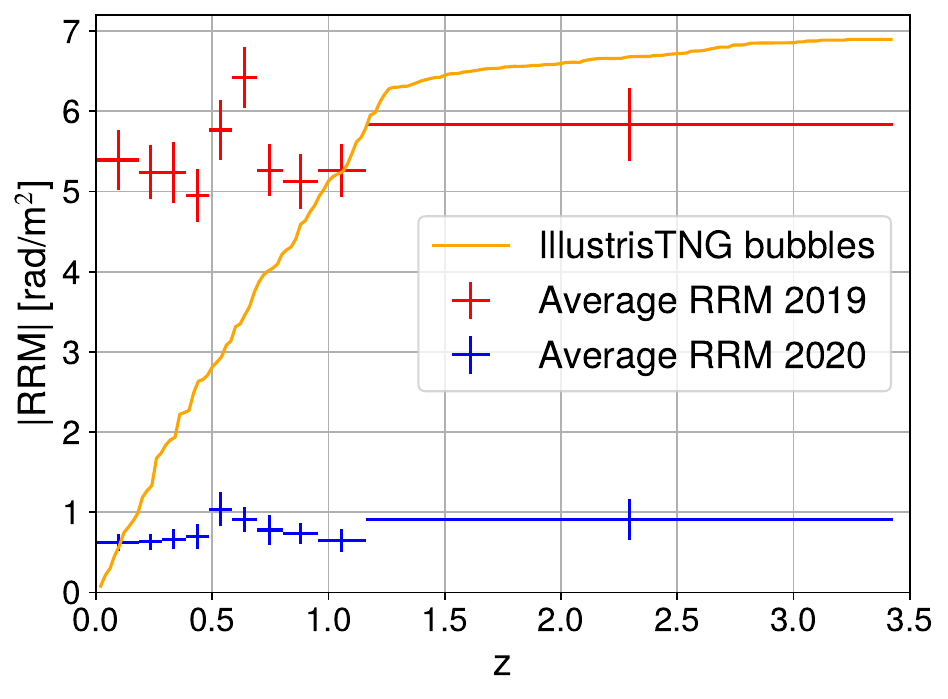}
    \caption{Average residual rotation measure (RRM) in different redshift bins for LoTSS survey. Red error bars show the average RRM for the galactic RM map of the year 2019~\cite{2020A&A...633A.150H}, while blue error bars are for the galactic RM map of the year 2020~\cite{2022A&A...657A..43H}. The orange line shows the average RM expected from overmagnetized bubbles in IllustrisTNG simulation~\cite{Aramburo-Garcia:2022gzn}.}
    \label{fig:LOFAR_RRM} 
\end{figure}

In paper~\cite{Aramburo-Garcia:2022ywn} it was shown that  (i) overmagnetized bubbles predicted by IllustrisTNG give a significant contribution to the extragalactic rotation measure (ii) this prediction is consistent with the data of the NVSS survey. In this section, we revise this analysis result using more accurate data from the LoTSS DR2 dataset.

\begin{figure}[h!]
    \centering
    \includegraphics[width=0.75\linewidth]{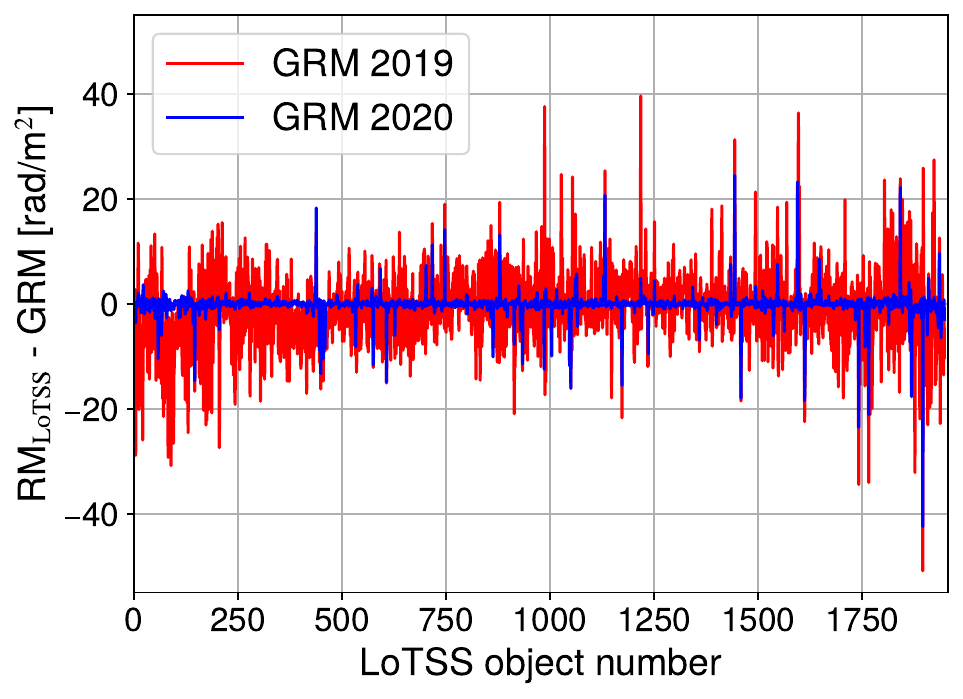}
    \caption{
    Residual rotation measure of LoTSS data for different observed objects. Red line corresponds to GRM from~\cite{2020A&A...633A.150H}, blue line corresponds to GRM from~\cite{2022A&A...657A..43H}.}
    \label{fig:RM_minus_GRM} 
\end{figure}

Using the LoTSS data we calculate the RRM using the two GRM maps described above and present results together with the contribution from magnetic bubbles in Fig.~\ref{fig:LOFAR_RRM}. We see that LoTSS RRM changes drastically between the two GRM maps.

To illustrate this difference, we also show in Fig.~\ref{fig:RM_minus_GRM} the RRM for each individual line of sight of the LoTSS data.  We see that the GRM 2020 map attributes a much smaller part of the total LoTSS RM to the extra-galactic contribution, as compared to the GRM 2019 map. 

Technically, the origin of this feature is that the GRM 2020 map, which was built using, in particular, the same LoTSS RM data, is much more correlated with those data (again, as compared to the GRM 2019 map). To illustrate this, we zoom in on two regions in the LoTSS map and compare the GRM 2020 and LoTSS data in more detail.

Specifically, we select two regions (see Appendix~\ref{sec:comparison-rrm}): region 1 with galactic coordinates $60^\circ<\ell<70^\circ$ and $40^\circ<b<50^\circ$ containing 30 objects, and region 2 with $110^\circ<\ell<120^\circ$ and $-40^\circ<b<-30^\circ$ containing 25 objects (see left panels of Figs.~\ref{fig:region1} and \ref{fig:region2} for visualization). Within each region, we sorted points in such a way that each point is connected to its closest neighbor (see right panels of Figs.~\ref{fig:region1} and \ref{fig:region2}). 

\begin{figure}
    \centering
    \includegraphics[width=0.47\textwidth]{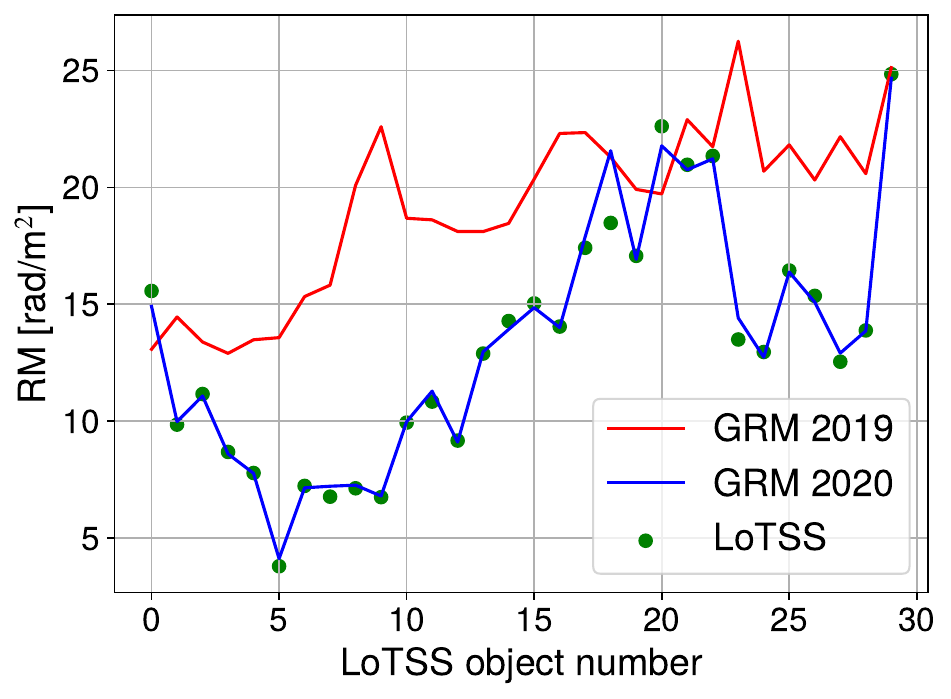}~\includegraphics[width=0.48\textwidth]{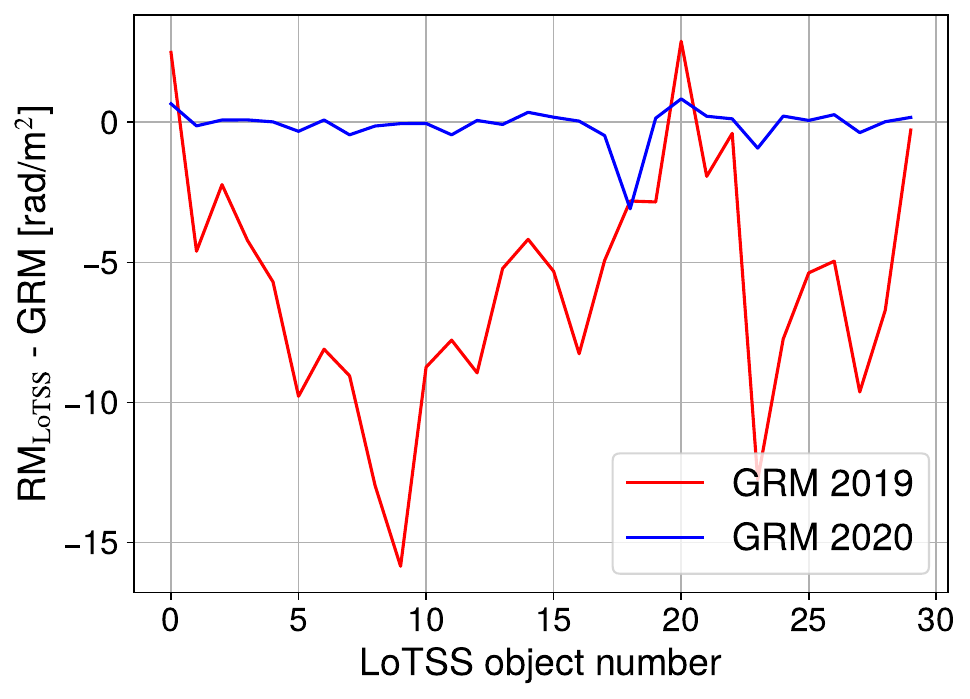}
    \caption{\textit{Left panel}: RM of LoTSS objects selected at Fig.~\ref{fig:region1} (green points). The red line shows galactic RM by the map GRM 2019 at the same positions as LoTSS points, while the blue line shows the same for the GRM 2020 map. \textit{Right panel}: residual RM for two GRM maps.}
    \label{fig:RRM-path1}
\end{figure}

\begin{figure}
    \centering
    \includegraphics[width=0.48\textwidth]{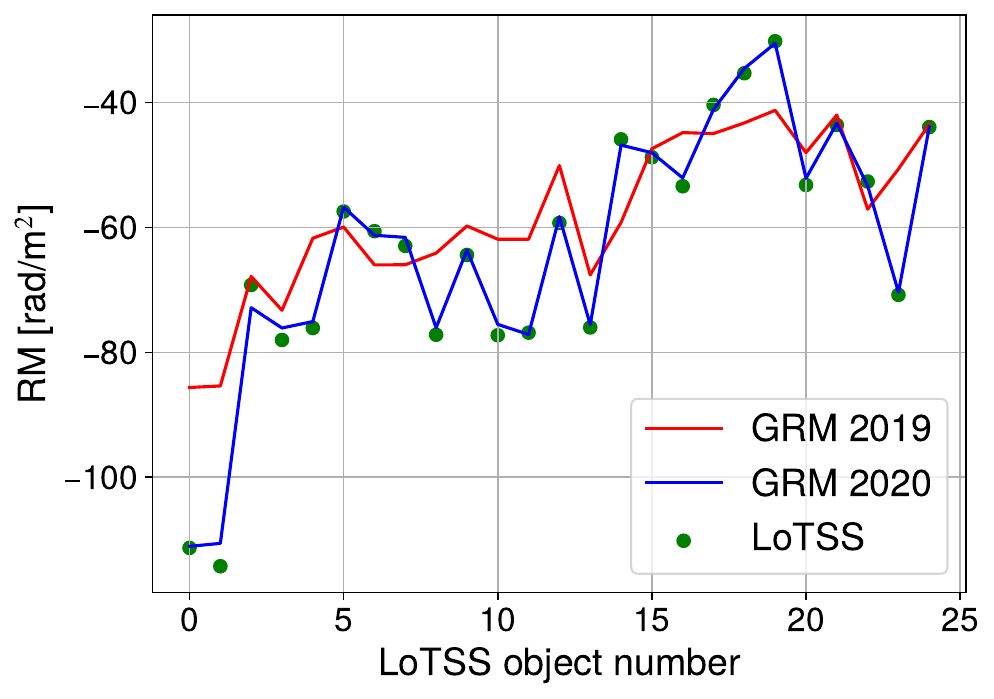}~\includegraphics[width=0.48\textwidth]{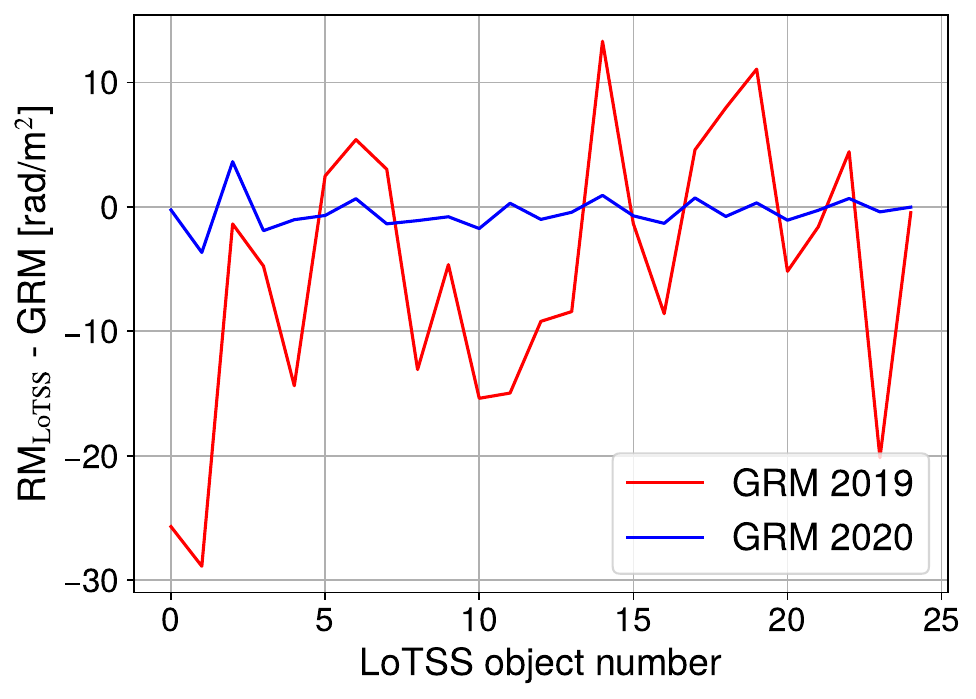}
    \caption{The same as Fig.~\ref{fig:RRM-path1}, but for the region from Fig.~\ref{fig:region2}.}
    \label{fig:RRM-path2}
\end{figure}

In Figs.~\ref{fig:RRM-path1} and~\ref{fig:RRM-path2} we compare
the GRM 2020, GRM 2019, and the total LoTSS RM from the two regions described above. We clearly see that GRM 2020 data  follows the LoTSS data much closer than the GRM 2019.
Taking into account the fact that (i) LoTSS data were used in building GRM 2020 and (ii) any local galactic data, like free free emission, was not used, we face a simple physical question -- having the total LoTSS RM along a given line of sight, what principle can we use to split it into galactic and extra-galactic contributions, to justify the difference that we see in  Figs.~\ref{fig:RRM-path1} and~\ref{fig:RRM-path2}? In~\cite{2022A&A...657A..43H} (GRM 2020) such split is done assuming that extragalactic contribution is not correlated for different objects and assuming Gaussian distribution of it. In this assumption, the Wiener filter allows to extract of the smooth component of the signal.

\begin{figure}[h!]
    \centering
    \includegraphics[width=0.75\linewidth]{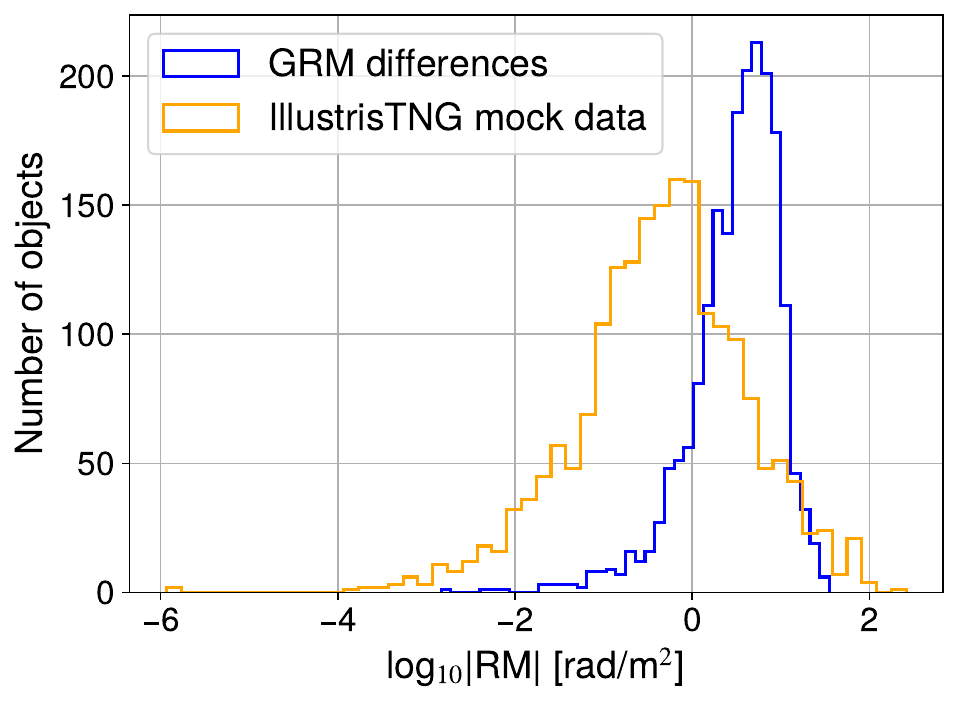}
    \caption{Blue histogram shows the distributions of the GRM difference between GRM 2019 and GRM 2020. The orange line shows RM from the mock IllustrisTNG data, where the redshifts and number of objects correspond to the observed LoTSS data.}
    \label{fig:deltaGRM_TNG} 
\end{figure}

We believe that, while GRM 2020 is a good estimate of the Galactic contribution on average, we can not use it to define extra-galactic contribution in the same LoTSS RM data, as this would mean that we know the splitting between Galactic and extra-galactic contribution quite precisely for each line of sight.  We do not think that the methodology of~\cite{2022A&A...657A..43H} is sufficient for such a claim and, therefore, in the current work, we interpret the difference between two GRM maps as a systematic error.  Therefore, to put conservative limits on the extra-galactic contribution from RM to LoTSS data, we will use the GRM 2019 map, which is more conservative in this aspect. In Fig.~\ref{fig:deltaGRM_TNG} we compare the contribution from over-magnetized bubbles predicted by IllustrisTNG with the systematic uncertainty in GRM is estimated as the difference between GRM 2020 and GRM 2019. We see that this contribution can be hidden within the uncertainty and therefore we conclude that currently IllustriseTNG model does not contradict the LoTSS observations.

\subsection{Constraints on primordial magnetic field}
\label{subsec:primordialMF}

\begin{figure}[h!]
    \centering
    \includegraphics[width=0.75\linewidth]{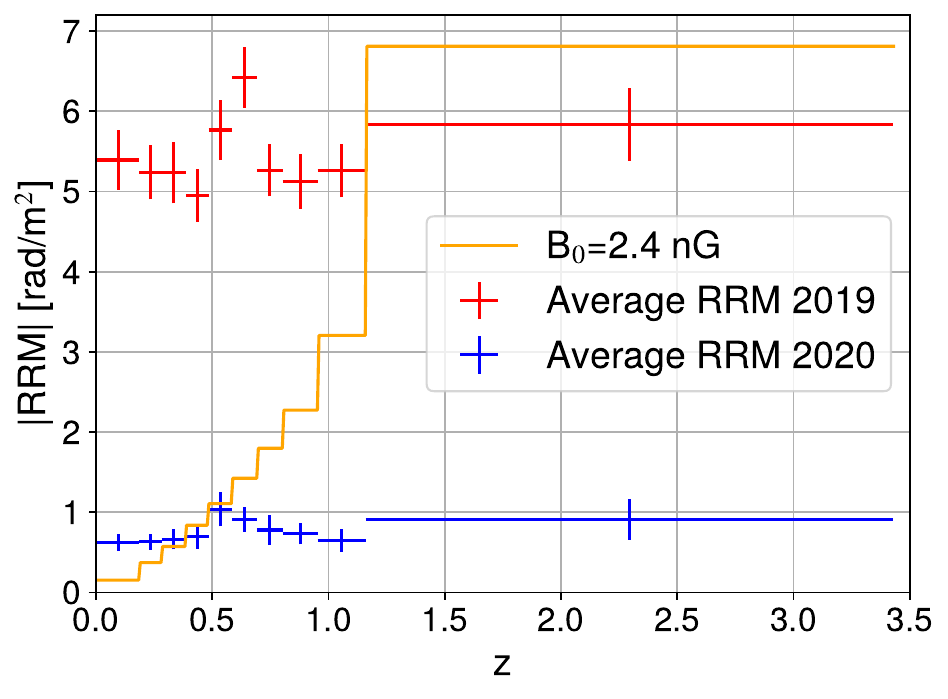}
    \caption{Residual rotation measure of LoTSS survey in different redshift bins. Red and blue error bars correspond to the galactic RM maps 2019~\cite{2020A&A...633A.150H} and 2020~\cite{2022A&A...657A..43H} correspondingly. The orange line shows the prediction for the average RM of the primordial magnetic field with strength $B_0 = 2.4$~nG from IllustrisTNG simulation.}
    \label{fig:PrimB_RRM} 
\end{figure}

Following work~\cite{Aramburo-Garcia:2022ywn} we divide simulation volume into regions where overmagnetized bubbles dominate and regions that are not significantly affected by galactic outflows. Specifically, we define the magnetic bubbles region by the condition $|B| > 10^{-12}$~cG, while all other simulation volume corresponds to the region where the contribution from the primordial magnetic field is the dominant one.

To put a conservative limit on the homogeneous primordial magnetic field (used in IllustrisTNG initial conditions), we generate 1000 continuous lines of sight up to redshift 5 and calculate rotation measure along each line of sight, excluding the regions with overmagnetized bubbles (see~\cite{Aramburo-Garcia:2022ywn} for details on generation continuous lines of sight from IllustrisTNG simulation box). We calculate average $|$RM$|$ as a function of redshift and multiply the result by the factor $B_0/10^{-14}$~cG, to emulate a larger value of the primordial magnetic field. This approach gives a conservative estimate of RM for the homogeneous primordial magnetic field, as for magnetic field values larger than $10^{-14}$~cG, which was used as the initial condition for IllustrisTNG simulation, we expect to have a large volume occupied by a primordial magnetic field.

The result is shown in Fig.~\ref{fig:PrimB_RRM}. Also, we show here the average $|$RRM$|$ calculated from LoTSS data using two recent galactic RM maps in different redshift bins. To put a conservative constraint, we use data from the GRM 2019 map and calculate $2\sigma$ exclusion limit using the $\chi^2$ method. We define
\begin{equation}
    \chi^2 = \sum_i \dfrac{({\rm RRM}_{\rm LoTSS} - {\rm RM}_{\rm TNG})^2}{(\Delta {\rm RRM}_{\rm LoTSS})^2},
\end{equation}
where we estimated $\Delta \text{RRM}_{\rm LoTSS}$ as a statistical error of average $|\text{RRM}|$ value within each bin. The sum is taken only in the bins where predicted RM$_{\rm TNG}$ is larger than the observed average $|\text{RRM}|$. This gives us an exclusion limit $B_0 = 2.4\cdot 10^{-9}$~cG, which is shown in Fig.~\ref{fig:PrimB_RRM}.

\section{Radio pairs}
\label{sec:radio-pairs}

\subsection{Radio pairs}

\begin{figure}[h!]
    \centering
    \includegraphics[width=0.48\linewidth]{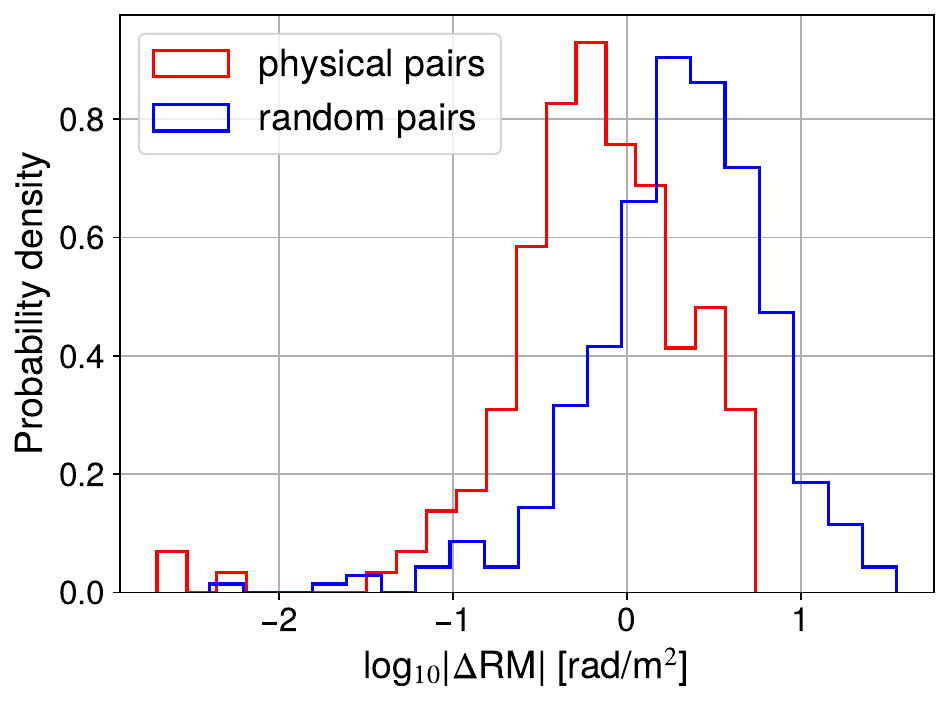}~\includegraphics[width=0.48\linewidth]{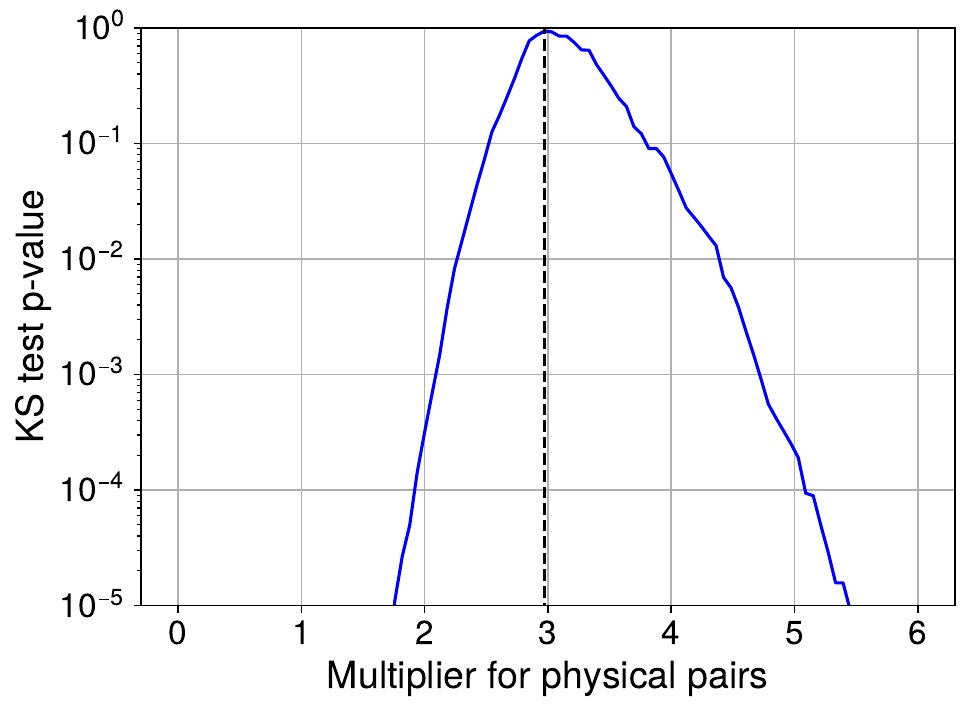}
    \caption{\textit{Left panel}: the distribution of $\Delta$RM distributions for physical (red) and random (blue) pairs. \textit{Right panel}: results of the Kolmogorov-Smirnov test between the distribution of random pairs and physical pair multiplied by a constant factor as a function of this factor.}
    \label{fig:RMpairsdistribution} 
\end{figure}

As we discuss in Sec.~\ref{subsec:rrm}, the procedure of accurate estimate of galactic RM and subtracting it to obtain an extragalactic one could give significant uncertainties. However, it is usually considered that GRM dominates over other components for most of the observed objects, so it is important to eliminate it for studying the extragalactic component. An alternative way to probe extragalactic RM is to use pairs of radio sources with a small angular separation and use the difference between their measured RMs, $\Delta$RM, as an observable. Indeed, we expect that the difference in RMs of the pair of close objects should cancel out the contribution from the GRM, as the line of sight of these pairs goes through a close region in the Galaxy. These pairs could be close to each other in the sky because they are two components from the same source (physical pairs) or they can come from different sources that are accidentally close to each other (random pairs). For the same reason as for the Galactic RM, for physical pairs, the contribution from the IGMF should approximately cancel out, while for the random pairs, there should be an IGMF contribution from the region between two sources (see~\cite{Pomakov:2022cem} for more details). An additional contribution to $\Delta$RM for both physical and random pairs could come from the local environment near the host galaxies.

For the LoTSS data, we follow methodology from~\cite{Pomakov:2022cem} and select close pairs with the angular separation $\Delta \theta < 0.5^{\circ}$. Next, we select physical and random pairs by the criteria that both components of the physical pair should have the same redshift. This gives us 169 physical pairs and 353 random pairs. We show the distribution of the RM differences for both samples in Fig.~\ref{fig:RMpairsdistribution} (left panel). Surprisingly, 
$|\Delta\text{RM}|$ distribution of physical and random pairs have the similar shape. To check this hypothesis we perform a Kolmogorov-Smirnov test between the distribution of random pairs and physical pairs multiplied by some factor $f$, see Fig.~\ref{fig:RMpairsdistribution} (right panel). It turns out, that two distributions are statistically indistinguishable for factor $f\approx 3$. This suggests us a simple method to subtract the fluctuation of RM measured in physical pairs from random pairs: we can introduce a correction factor $1 - 1/f \approx 0.66$ to the random pairs. This procedure allows us to estimate the contribution from IGM between redshifts $z_1$ and $z_2$ of random objects more conservatively.

\begin{figure}
    \centering
    \includegraphics[width=0.48\textwidth]{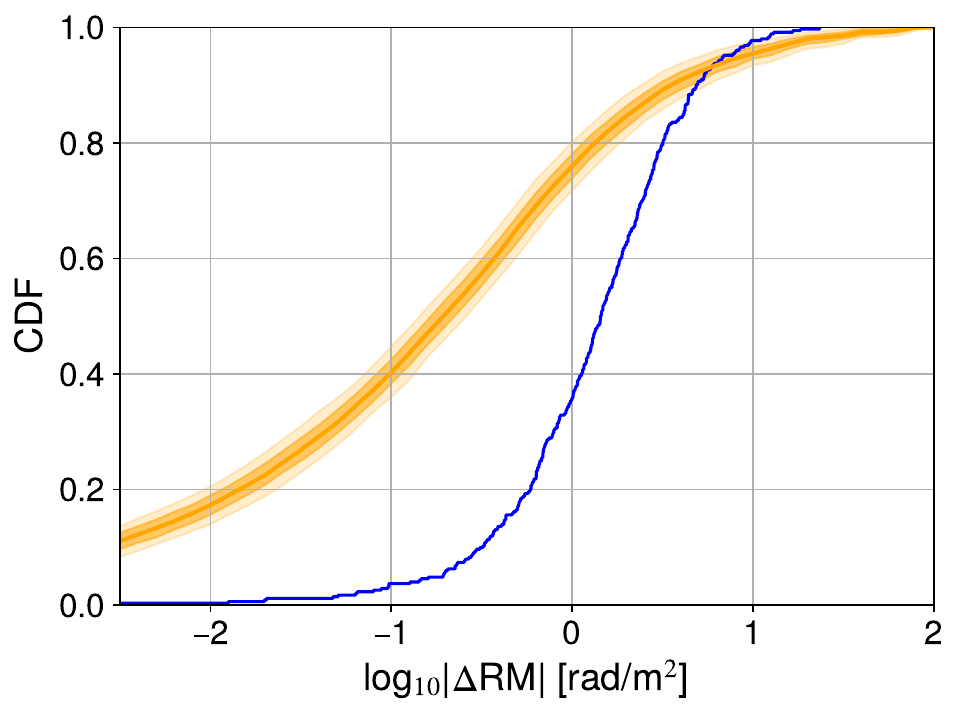}~\includegraphics[width=0.48\textwidth]{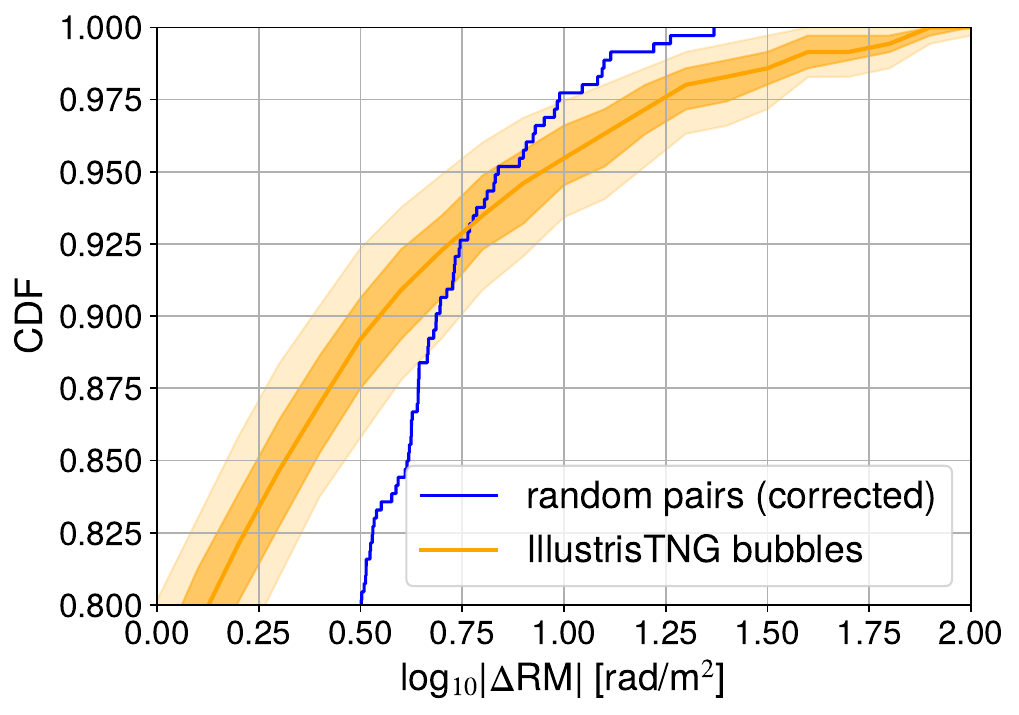}
    \caption{\textit{Left panel}: The blue line shows the cumulative distribution function of $|\Delta$RM$|$ distribution for random pairs with correction that takes into account $|\Delta$RM$|$ in physical pairs (see text for details). The orange line and regions show a contribution from IGM with magnetic bubbles calculated with IllustrisTNG simulation. The central orange line corresponds to the median expected contribution, darker and lighter orange regions show $68\%$ and $95\%$ sample variance in simulations. \textit{Right panel}: zoom-in to the region of large $|\Delta$RM$|$.}
    \label{fig:CDF-pairs}
\end{figure}

\subsection{Radio pairs and magnetic bubbles}

Using the method of close radio pairs, we would like to test the feedback model in the IllustrisTNG simulation. To do this we prepare a mock data set in the simulation that corresponds to the observed set of random pairs. Specifically, for each observed pair, we take a random line of sight from the IllustrisTNG simulation and calculate $\Delta$RM between the redshifts $z_1$ and $z_2$ of two sources in this random pair. This gives us a prediction for the contribution from the IGMF for the IllustrisTNG model between the two sources. We repeat this procedure 1000 times, creating many mock realizations of observed pairs. 

To visualize our results we use the empirical cumulative distribution function, which is defined as
\begin{equation}
    \text{CDF}(|\Delta \text{RM}|) = 
    \frac{1}{N} \sum_{i} \theta(|\Delta \text{RM}| - |\Delta \text{RM}_i|),
\end{equation}
where $\theta(x)$ is a unit step function and $N$ is a number of pairs. We show empirical cumulative distribution functions for mock data and observed random pairs with a correction factor $0.66$ introduced in the previous section in Fig.~\ref{fig:CDF-pairs}. We see that for $90\%$ of pairs, the $\Delta$RM contribution from IGM in simulation is small or even much smaller than the observed ones. However, for high $\Delta$RM tails of distributions, we found that IllustrisTNG often predicts some lines of sight with the value of $\Delta$RM larger than observed in LoTSS. Only for $0.5\%$ of our random realizations of the observed pair, the contributions predicted by IllustrisTNG are smaller than observed $\Delta$RM. This tension comes from pairs with the largest difference of redshift, see Appendix~\ref{sec:pairs-bins} for more details.

\begin{figure}[h!]
    \centering
    \includegraphics[width=0.48\linewidth]{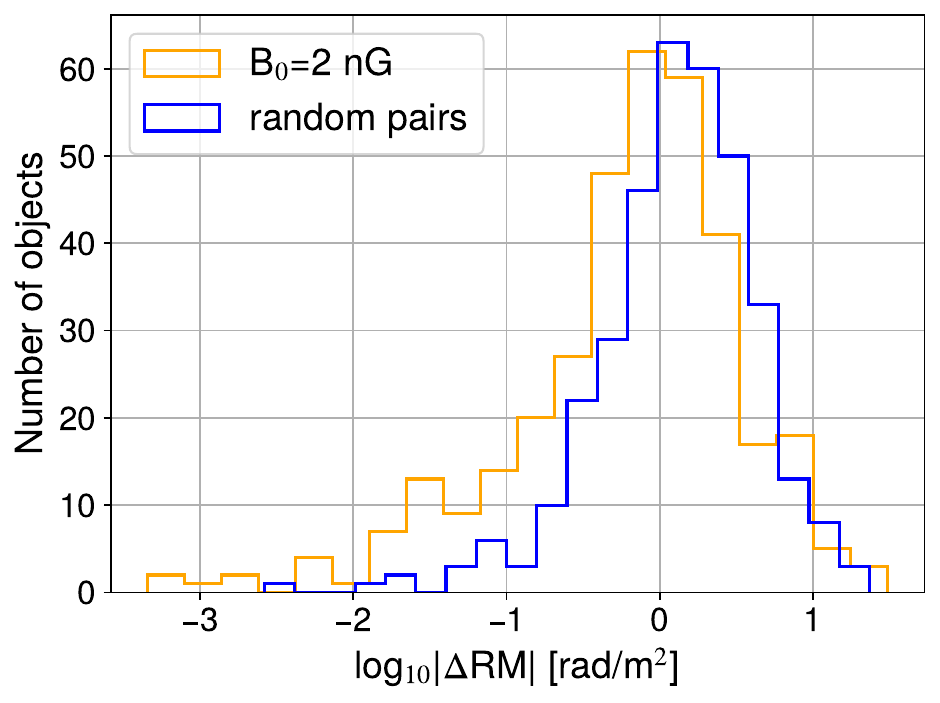}~\includegraphics[width=0.48\linewidth]{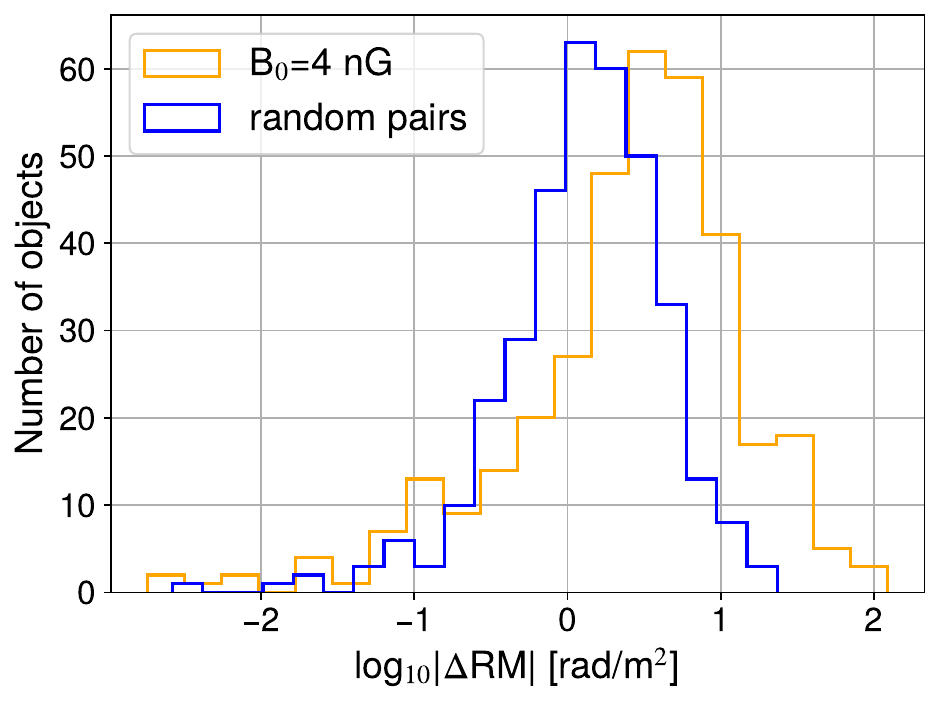}
    \caption{Blue line shows the distribution of $|\Delta\text{RM}|$ for random pairs with the correction that takes into account $|\Delta$RM$|$ observed for physical pairs (see text for details).
    The orange line shows our estimation for the contribution to $|\Delta\text{RM}|$ from the primordial homogeneous magnetic field from IllustrisTNG simulation with the magnetic field strength $2 \cdot 10^{-9}$~cG (left panel) and  $4 \cdot 10^{-9}$~cG (right panel).}
    \label{fig:TNG_B_constraints} 
\end{figure}

\subsection{Radio pairs and constraints on primordial magnetic field}

In this section, we test the method of close radio pairs for putting limits on the primordial homogeneous magnetic field. We use the same methodology as in Sec.~\ref{subsec:primordialMF} to generate 1000 random lines of sight between redshifts $z=0$ to $z=5$ that do not contain a contribution from magnetic bubbles and scale the resulting RM by the factor $B_0/10^{-14}$~cG.

To model the contribution from primordial magnetic field to the random pairs, we calculate $\Delta$RM in the same way as in the previous section and show examples of the resulting distributions in Fig.~\ref{fig:TNG_B_constraints}, where we show results for the magnetic field values $B_0 = 2 \cdot 10^{-9}$~cG and  $B_0 = 4 \cdot 10^{-9}$~cG. We see that distribution with $B_0 = 2 \cdot 10^{-9}$~cG does not contradict the data, while for $B_0 = 4 \cdot 10^{-9}$~cG the high-RM tail of the distribution clearly contradicts observations. 

We conclude that the method of close radio pairs allows us to put a constraint at the level of $2-4$~nG, which is weaker than the previous result from NVSS data.  The reason for this is that in the NVSS data, constraints mainly come from large redshifts ($z \sim 3 - 5$). The LoTSS observations are mostly concentrated on small redshifts ($z \lesssim 2$) with a few objects up to $z \sim 3.5$ which does not allow to obtain a significant contribution from the primordial magnetic field.

\section{Conclusions}
\label{sec:conclusions}

In this work, we used the latest RM catalog LoTSS DR2 from LOFAR to probe the extragalactic cosmic magnetic fields in order to (i) test the prediction of the presence of Mpc-scale overmagnetized bubbles by the IllustrisTNG simulation~\cite{Garcia:2020kxm}, and (ii) derive limits on the presence of a volume-filling magnetic field with homogeneous initial conditions. To probe the extragalactic component of the observed RM, we use two different techniques: residual rotation measure and close radio pairs.

As we show in Sec.~\ref{subsec:rrm}, for the method of residual rotation measure (RRM) our results significantly depend on the version of the Galactic rotation measure (GRM) map used.  Fig.~\ref{fig:LOFAR_RRM} indicates that an older galactic map (GRM 2019~\cite{2020A&A...633A.150H}) gives systematically higher RRM than a latter one (GRM 2020~\cite{2022A&A...657A..43H}). We observe, that while both maps used a similar amount of extragalactic objects, to build the GRM 2020 map the same LoTSS DR2 RM catalog was used, and, as a result, this map repeats variations of RM observed in LoTSS DR2 at $1^\circ$ scales quite precisely. The GRM 2019 map did not use the LOFAR data and correlated its results with the map of free free emission data. Therefore, we conservatively estimate the systematic errors in GRM as the difference between two maps which does not allow us to exclude the presence of a magnetized bubble predicted by IllustrisTNG simulation. 

To put a constraint on a homogeneous volume-filling magnetic field we conservatively use a less constraining GRM 2019 map, with results in constraint at the level of $2.4$~nG, which is slightly weaker than in the previous analysis of NVSS data~\cite{Aramburo-Garcia:2022ywn}. LoTSS RM measurements have much higher precision than the NVSS survey. The reason for the slightly weaker sensitivity of LoTSS data is smaller redshift coverage (up to $z\approx3.5$) compared to the NVSS (up to $z\approx 5$), while expected RM contribution from volume-filling magnetic field quickly grows with redshift, see e.g. Fig.~\ref{fig:PrimB_RRM}.

An alternative approach to the analysis of the LoTSS RM catalog is the study of differences between rotation measures of close radio pairs. Using this method, one could expect to cancel out the contribution from our Galaxy with good precision as well as a part of the intergalactic medium that is common for both lines of side, leaving mostly the IGM contribution from the part of the longer LOS between the sources and contribution from the local regions near the sources. Applying this method to the  LoTSS data and comparing it with the mock data from the IllustrisTNG simulations, we found that only in 0.5\% of our mock realizations of observed data the expected contribution from the over-magnetized bubbles does not exceed LoTSS DR2 data. 
% we found that in only XXX of our mock realizations of observed data from YYY the expected contribution from the over-magnetized bubbles does not exceed LoTSS DR2 data.  
We have also shown, that constraints on the presence of a volume-filling homogeneous magnetic field are at the level of $(2\div 4)\cdot 10^{-9}$~cG, which is slightly weaker than the results of our RRM study. 

Our study indicates, that to advance our probes of intergalactic magnetic field one needs not only to have a good precision of RM measurement for individual objects but also to (a) develop a more accurate method of subtraction of galactic components; (b) perform a survey with larger depth, covering larger redshift range; (c) make a survey with higher density of sources, which will increase the probability to form random pair with a significant redshift differences between objects.

\section*{Acknowledgements}
We acknowledge support from {\sl Fermi Research Alliance, LLC} under Contract No. DE-AC02-07CH11359 with the U.S. Department of Energy, Office of High Energy Physics. Nordita is supported in part by NordForsk.

\appendix

\section{Comparison of RRM between two GRM maps}
\label{sec:comparison-rrm}

\begin{figure}
    \centering
    \includegraphics[width=0.48\textwidth]{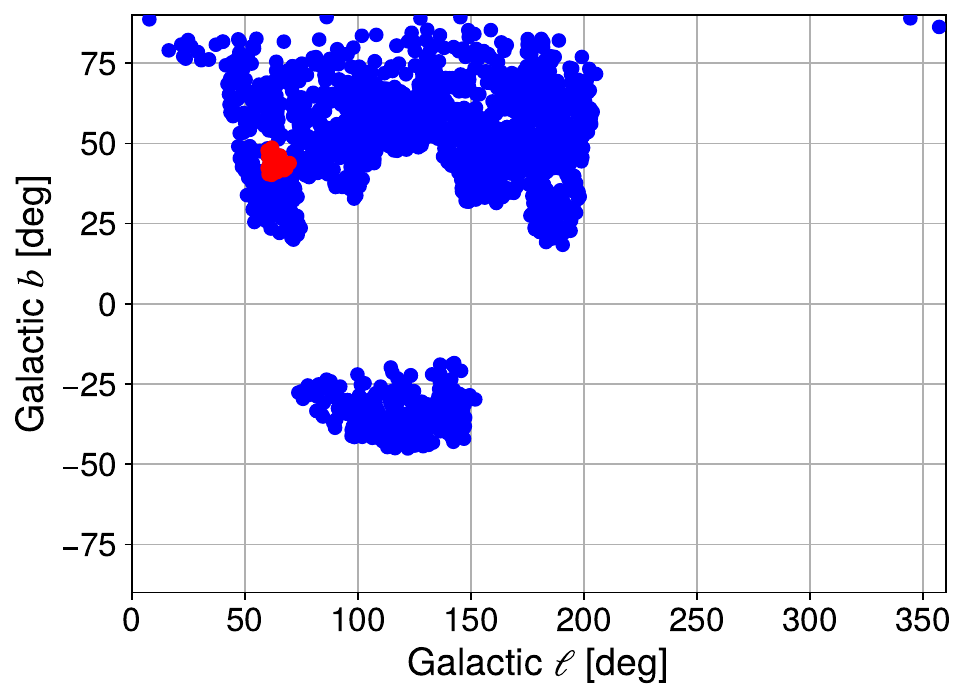}~\includegraphics[width=0.48\textwidth]{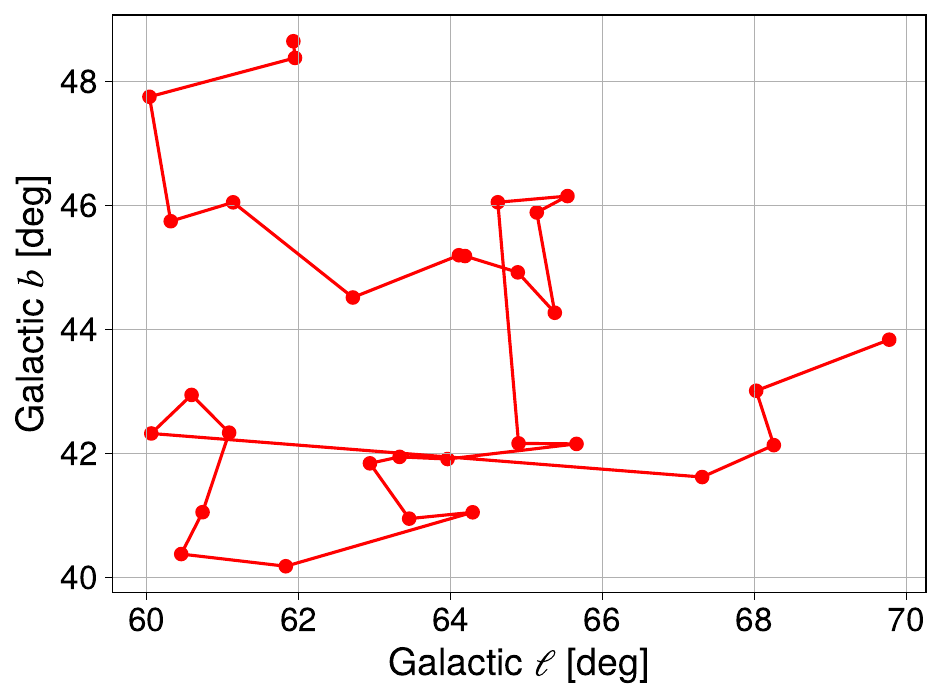}
    \caption{\textit{Left panel}: positions on the galactic plane of 30 points selected within region $60^\circ<\ell<70^\circ$ and $40^\circ<b<50^\circ$ (red points) on the background of all LoTSS measurement (blue points). \textit{Right panel}: zoom-in to the selected region. The red line illustrates the order of points that we used in Fig.~\ref{fig:RRM-path1}.}
    \label{fig:region1}
\end{figure}

\begin{figure}
    \centering
    \includegraphics[width=0.48\textwidth]{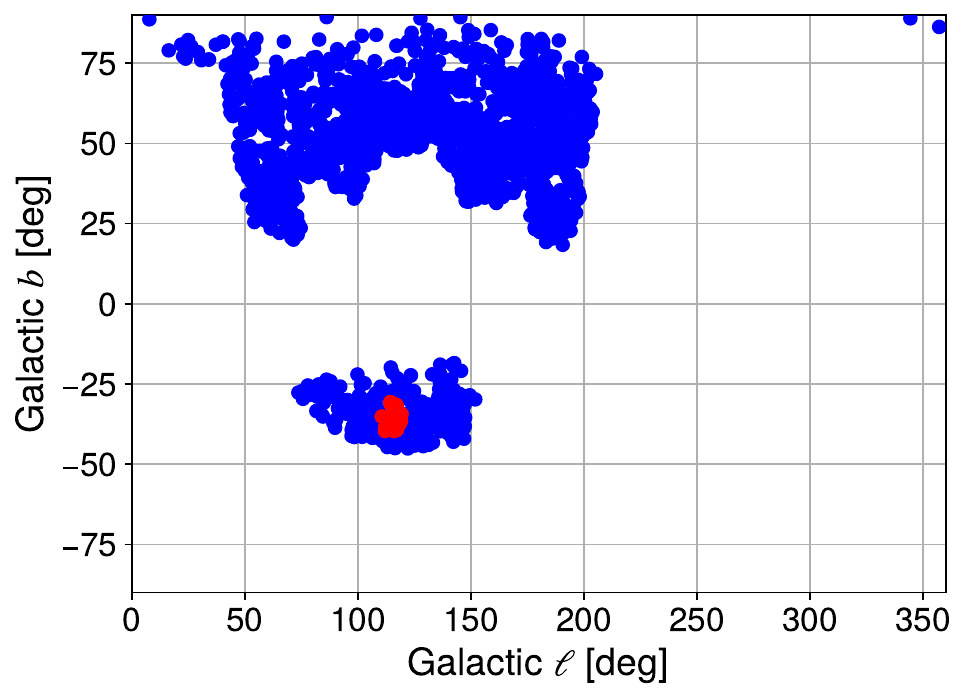}~\includegraphics[width=0.48\textwidth]{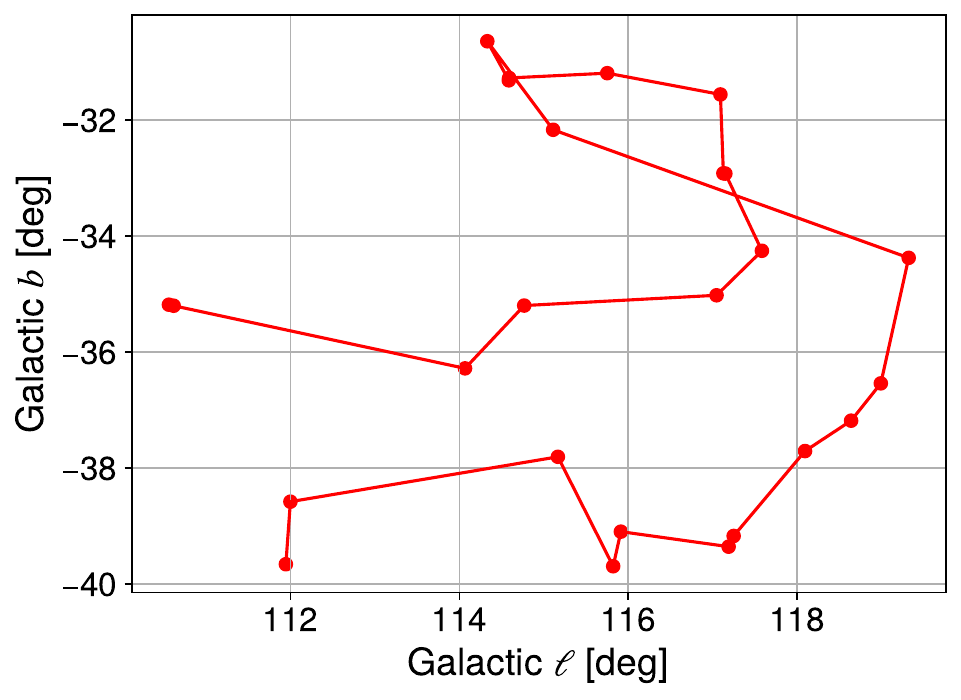}
    \caption{The same as Fig.~\ref{fig:region1}, but for the sky region $110^\circ<\ell<120^\circ$ and $-40^\circ<b<-30^\circ$ with 25 points in it for Fig.~\ref{fig:RRM-path2}.}
    \label{fig:region2}
\end{figure}

In this appendix, we show how we select objects for Figs.~\ref{fig:RRM-path1} and \ref{fig:RRM-path2} in the main text. We choose two regions: region 1 with galactic coordinates $60^\circ<\ell<70^\circ$ and $40^\circ<b<50^\circ$ containing 30 objects, and region 2 with $110^\circ<\ell<120^\circ$ and $-40^\circ<b<-30^\circ$ containing 25 objects (see left panels of Figs.~\ref{fig:region1} and \ref{fig:region2} for visualization). Within each region, we sorted points in such a way that each point is connected to its closest neighbor (see right panels of Figs.~\ref{fig:region1} and \ref{fig:region2}). 

\section{Studying close radio pairs in different redshift bins}
\label{sec:pairs-bins}

\begin{figure}
    \centering
    \includegraphics[width=0.75\textwidth]{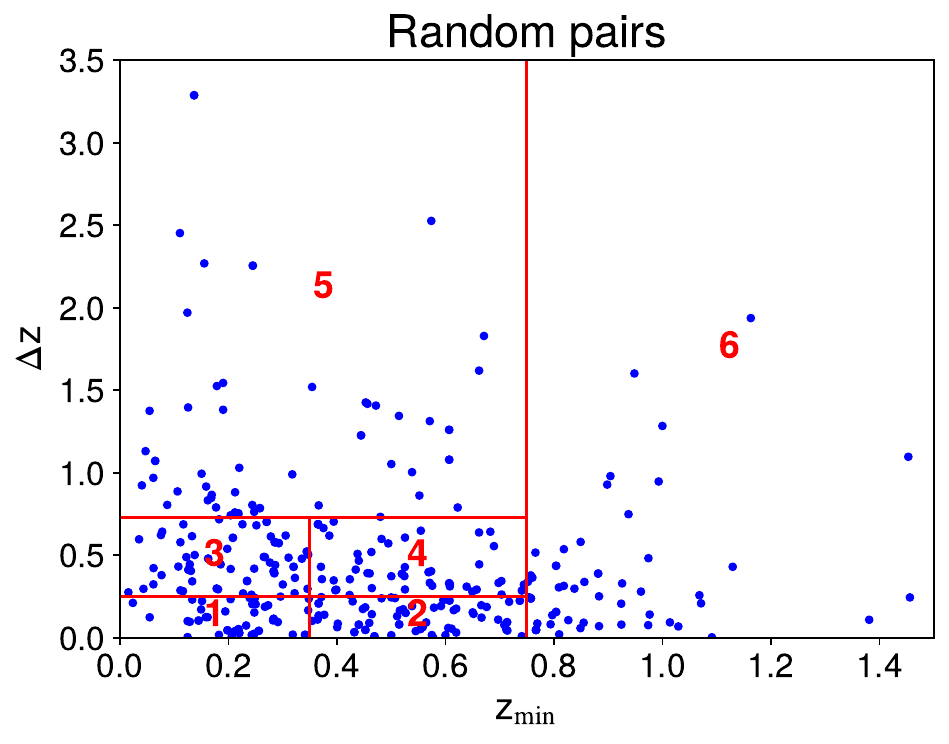}
    \caption{Blue point shows distribution of random close radio pairs from LoTSS survey by minimal redshift of objects in pair (x-axis), and by redshift difference between objects in pair (y-axis). Red lines indicate our selection of bins in this 2D parameter region, see main text for details.}
    \label{fig:pairs-z-bins}
\end{figure}

Each random pair contains two objects with redshifts $z_1$ and $z_2$. We can describe properties of a random pair in redshift space with minimal redshift of the pair $z_{\min} = \min(z_1,z_2)$ and redshift difference between objects in the pairs, $\Delta z = |z_1 - z_2|$. The distribution of random pairs from LoTSS in this two-dimensional space is shown in Fig.~\ref{fig:pairs-z-bins}. To study pairs in more detail we introduce six bins in this two-dimensional space, with approximately $60$ objects in each bin. Details of bin selection are given in Tab.~\ref{tab:pair-bins} and are visualized in Fig.~\ref{fig:pairs-z-bins}.

\begin{figure}
    \centering
    \includegraphics[width=0.48\textwidth]{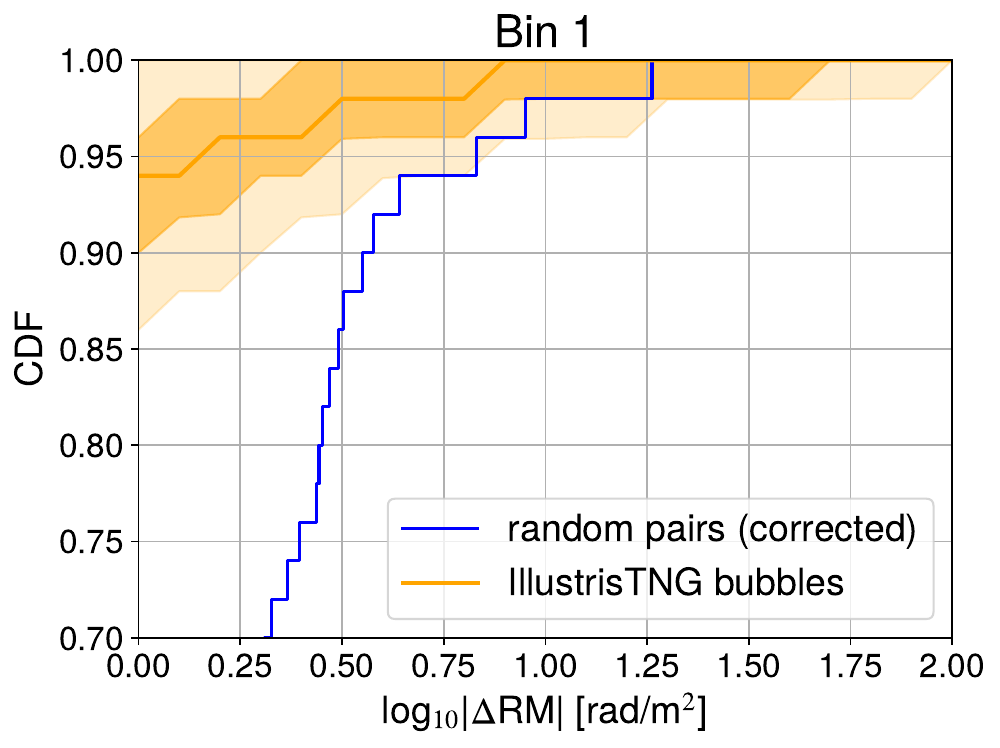}~\includegraphics[width=0.48\textwidth]{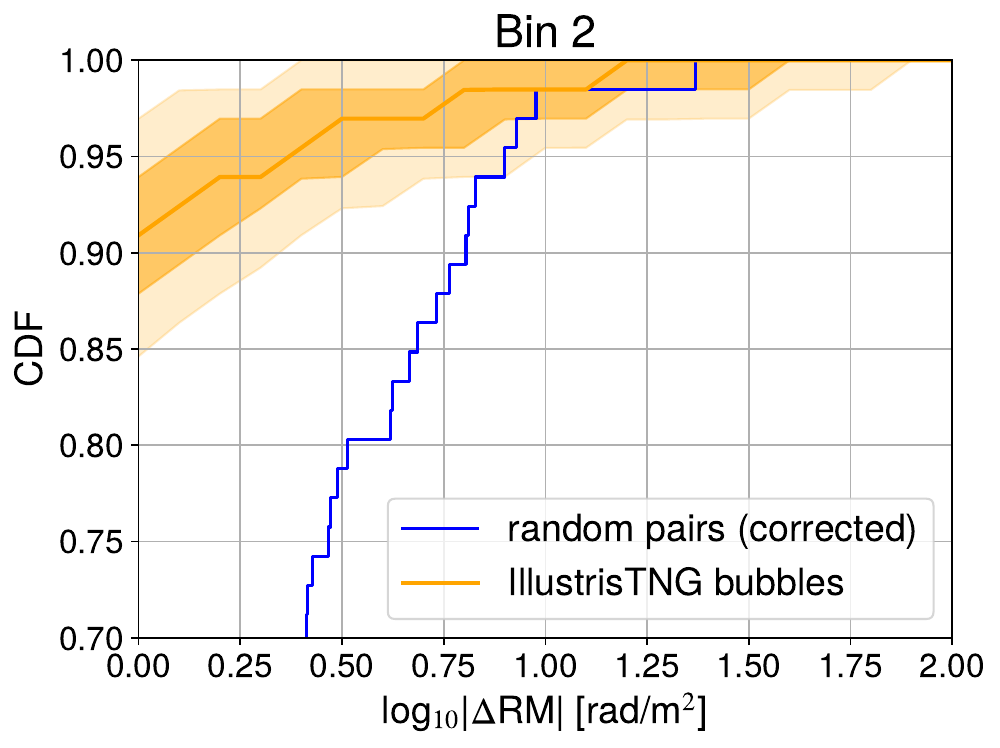}
    \\
    \includegraphics[width=0.48\textwidth]{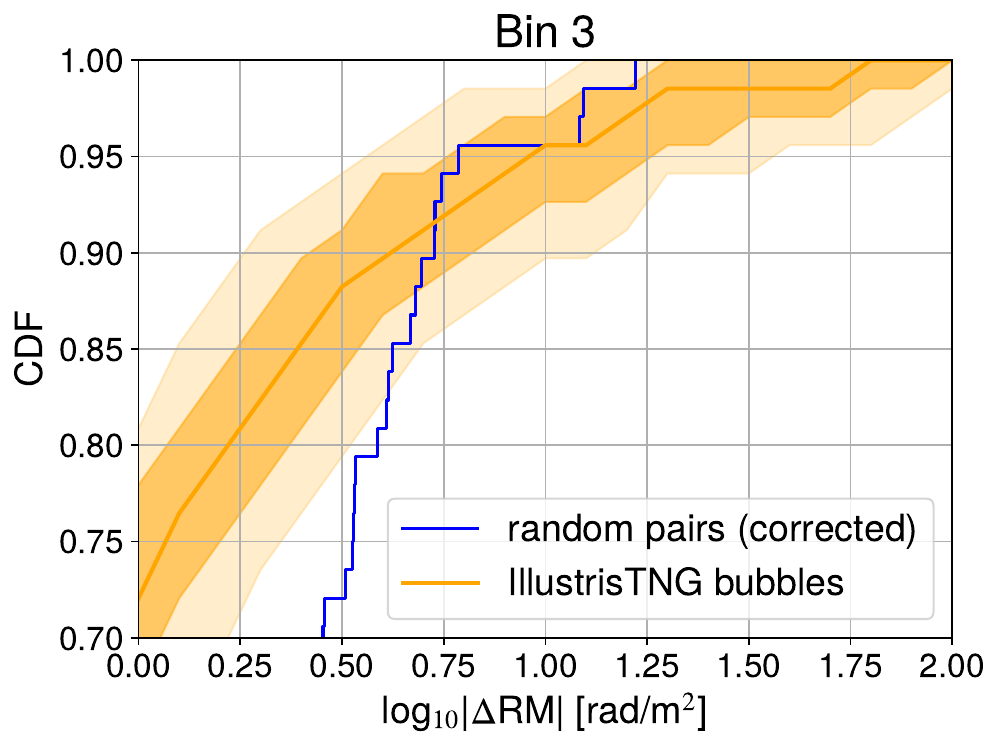}~\includegraphics[width=0.48\textwidth]{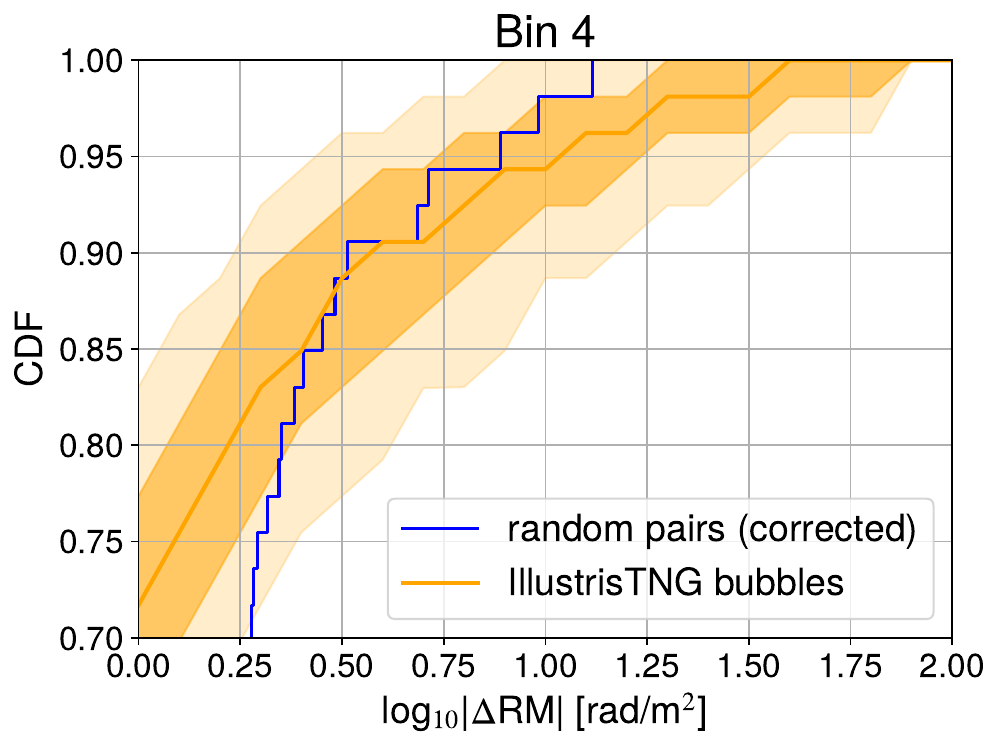}
    \\
    \includegraphics[width=0.48\textwidth]{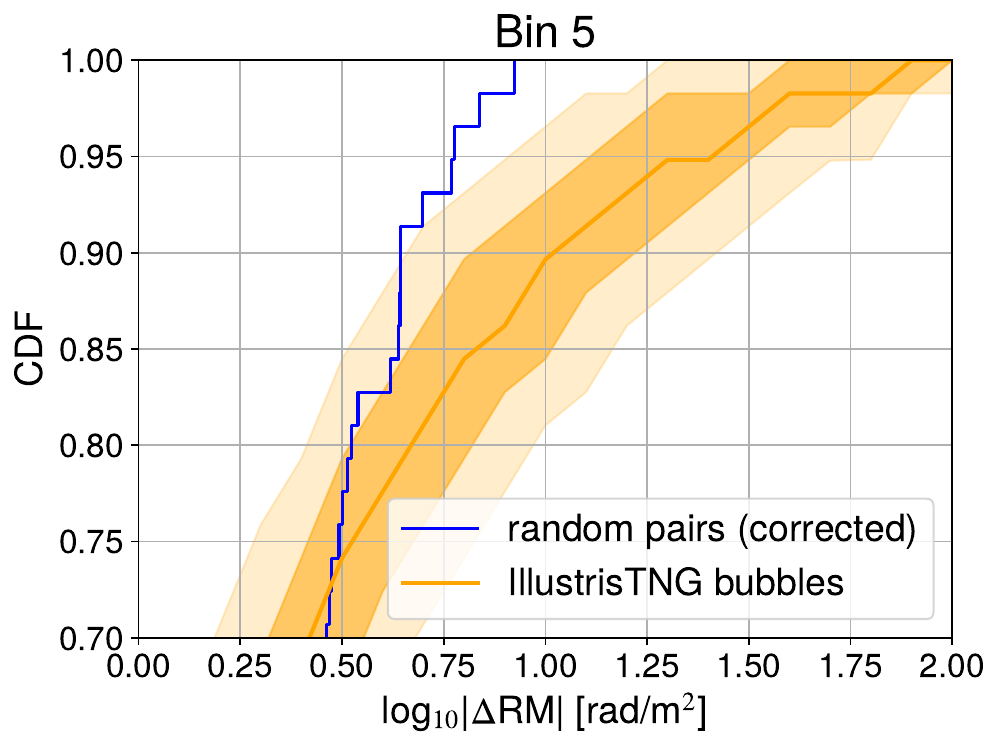}~\includegraphics[width=0.48\textwidth]{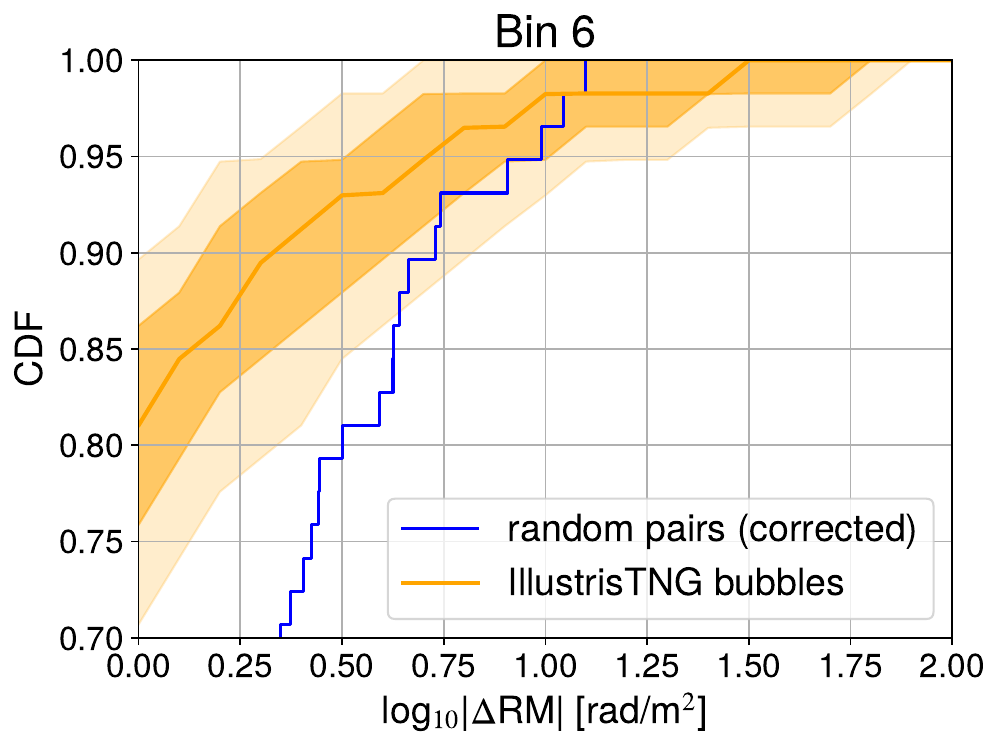}
    \caption{The same as Fig.~\ref{fig:CDF-pairs} in different bins, see binning in Fig.~\ref{fig:pairs-z-bins}. Mock data (orange regions and line) is also generated only for the objects in the corresponding bin.}
    \label{fig:CDF-z-bins}
\end{figure}

In Fig.~\ref{fig:CDF-z-bins} we show the empirical cumulative distribution function of $|\Delta$RM$|$ for pairs from different redshift bins and compare them to the mock pairs generated from the IllustrisTNG simulation in the same way as in the main text. We see that the only bin that has significant tension between simulations and observations is bin 5, which corresponds to objects with small $z_{\min}$ but large $\Delta z$. These are expected results, as a contribution to RM from IGM in simulations grows with $\Delta z$. Detail studies of this bin show, that only in $0.5\%$ of our randomly generated mock data there are no lines of sight that have larger RM than observed $\Delta$RM.

\begin{table}[]
\centering
\begin{tabular}{|l|c|c|c|c|c|c|}
\hline
Bin number             & 1        & 2           & 3           & 4           & 5          & 6          \\ \hline
$z_{\text{min}}$ range & $0\div0.35$ & $0.35\div0.75$ & $0\div0.35$    & $0.35\div0.75$ & $0\div0.75$   & $0.75\div1.5$ \\ \hline
$\Delta z$ range       & $0\div0.25$ & $0\div0.25$    & $0.25\div0.73$ & $0.25\div0.73$ & $0.73\div3.5$ & $0\div3.5$    \\ \hline
Pairs num.       & 50       & 66          & 68          & 53          & 58         & 58         \\ \hline
\end{tabular}
\caption{The redshift bins for physical pairs, where $z_{\min}$ is the minimal redshift of an object in a pair, and $\Delta z$ is the redshift difference between the pair. }
\label{tab:pair-bins}
\end{table}

\bibliographystyle{JHEP}
\bibliography{refs.bib}

\end{document}